\documentclass{amsart}%
\usepackage{graphicx}
\usepackage{amsmath}
\usepackage{amsfonts}
\usepackage{amssymb}%
\begin{document}
\title{Two and Three Qubits Geometry and Hopf Fibrations }
\author{Remy Mosseri }
\address{Groupe de Physique des Solides, CNRS UMR 7588,\\
Universit\'{e}s Pierre et Marie Curie Paris 6 et Denis Diderot Paris 7,\\
2 place Jussieu, 75251 Paris Cedex 05 France\\
 }
\email{mosseri@ccr.jussieu.fr; }
\maketitle

\begin{abstract}
This paper reviews recent attempts to describe the two- and three-qubit
Hilbert space geometries with the help of Hopf fibrations. In both cases, it
is shown that the associated Hopf map is strongly sensitive to states
entanglement content. In the two-qubit case, a generalization of the one-qubit
celebrated Bloch sphere representation is described.

\end{abstract}

\section{ Introduction}

Two-level quantum systems, denoted qubits, have gained a renewed interest in
the past ten years, owing to the fascinating perspectives of quantum
information\cite{q-info-gen}. Having in mind the different qubit manipulation
protocols that are proposed in this growing field, it is therefore of high
interest to represent their quantum evolution in a suitable representation
space, in order to get some insight into the subtleties of this complicated
problem. For single two-level systems, a well known tool in quantum optics is
the Bloch sphere representation, where the simple qubit state is faithfully
represented, up to a global phase, by a point on a standard sphere $S^{2}$,
whose coordinates are expectation values of physically interesting operators
for the given quantum state. Guided by the relation between the Bloch sphere
and a geometric object called the Hopf fibration of the $S^{3\text{ }}%
$hypersphere\cite{urtbanke}, a generalization for a two-qubit system was
recently proposed\cite{mosseridandoloff}, in the framework of the (high
dimensional) $S^{7}$ sphere Hopf fibration, and will be recalled below. An
interesting result is that the $S^{7}$ Hopf fibration is entanglement
sensitive and therefore provides a kind of \textquotedblright
stratification\textquotedblright\ \ for the 2 qubits states space with respect
to their entanglement content. An extension of this description to a three
qubits system, using the $S^{15}$ Hopf fibration, will also be presented here.

We first briefly remind known facts about the Bloch sphere representation, and
its close relation to the $S^{3}$ Hopf fibration. We then recall in some
details what was recently done for the two-qubit case in terms of the $S^{7}$
Hopf fibration. The $S^{15}$ Hopf fibration is then introduced, which helps
describing the three-qubits Hilbert space geometry. As far as computation is
concerned, going from the $S^{3}$ to the $S^{7}$ and then the $S^{15}$
fibrations merely amounts to replacing complex numbers by quaternions and then
octonions. This is why a brief introduction to quaternions and octonions is
given in appendix. Note that using these two kinds of generalized numbers is
not strictly necessary here, but they provide an elegant way to put the
calculations into a compact form, and have (by nature) a natural geometrical interpretation

\section{From the S$^{3}$ hypersphere to the Bloch sphere representation}

A (single) qubit state reads
\begin{equation}
\left\vert \Psi\right\rangle =\alpha\left\vert 0\right\rangle +\beta\left\vert
1\right\rangle ,\qquad\alpha,\beta\in\mathbb{C}\mathbf{,\qquad}\left\vert
\alpha\right\vert ^{2}+\left\vert \beta\right\vert ^{2}=1
\label{one-qubit state}%
\end{equation}

In the spin
${\frac12}$
context, the orthonormal basis $\left\{  \left\vert 0\right\rangle ,\left\vert
1\right\rangle \right\}  $ are the two eigenvectors of the (say) $\sigma_{z}$
(Pauli spin) operator. Viewed as pairs of real numbers, the two normalized
components $\alpha,\beta$ generate a unit radius sphere S$^{3}$ embedded in
\textbf{R}$^{4}$. To take into account the global phase freedom, one expects
to find a way to fill S$^{3}$ with circles (the orbit of a global phase $\exp
i\omega$ multiplying the pair ($\alpha,\beta$)), such that each state belongs
to exactly one such circle. This task is nicely fulfilled by the so-called
S$^{3}$ Hopf fibration \cite{hopf}.

A fibred space $E$ is defined by a (many-to-one) map from $E$ to the so-called
\textquotedblright base space\textquotedblright, all points of a given fibre
$F$ being mapped onto a single base point. A fibration is said "trivial" if
the base $B$ can be embedded in the fibred space $E$, the latter being
faithfully described as the direct product of the base and the fibre (think
for instance of fibrations of $R^{3}$ by parallel lines $R$ and base $R^{2}$
or by parallel planes $R^{2}$ and base $R$).

The simplest, and most famous, example of a non trivial fibration is the Hopf
fibration of $S^{3}$ by great circles $S^{1}$ and base space $S^{2}$. For the
qubit Hilbert space purpose, the fibre represents the global phase degree of
freedom, and the base $S^{2}$ is identified to the Bloch sphere. One standard
notation for a fibred space is that of a map $E\overset{F}{\rightarrow}B$,
which reads here $S^{3}\overset{S^{1}}{\rightarrow}S^{2}$. Its non trivial
character implies $S^{3}\neq S^{2}\times S^{1}$. This translates into the
known failure in ascribing consistantly a definite phase to each representing
point on the Bloch sphere.

To describe this fibration in an analytical form, we go back to the definition
of $S^{3}$ as pairs of complex numbers $\left(  \alpha,\beta\right)  $ which
satisfy $\left\vert \alpha\right\vert ^{2}+\left\vert \beta\right\vert ^{2}%
=1$. The Hopf map is defined as the composition of a map $h_{1}$ from $S^{3}$
to $R^{2}$ $(+\infty)$, followed by an inverse stereographic map $h_{2\text{
}}$from $R^{2}$ to $S^{2}:$%

\begin{align}
h_{1}  &  :%
\begin{array}
[c]{ccc}%
S^{3} & \longrightarrow & R^{2}+\left\{  \infty\right\} \\
\left(  \alpha,\beta\right)  & \longrightarrow & C=\overline{\alpha\beta^{-1}}%
\end{array}
\qquad\alpha,\beta\in\mathbb{C}\nonumber\\
h_{2}  &  :%
\begin{array}
[c]{ccc}%
R^{2}+\left\{  \infty\right\}  & \longrightarrow & S^{2}\\
C & \longrightarrow & M(X,Y,Z)
\end{array}
\qquad X^{2}+Y^{2}+Z^{2}=1
\end{align}

where $\overline{z}$ is the complex conjugate of $z$). The first map $h_{1}$
clearly shows that the full $S^{3}$ great circle, parametrized by ($\alpha\exp
i\omega,\beta\exp i\omega$), is mapped onto the same single point with complex
coordinate $C$. It is easy to show that, with $R^{2}$ cutting the unit radius
$S^{2}$ along the equator, and the north pole (along the $Z$ axis) as the
stereographic projection pole, the $S^{2}$ Hopf fibration base coordinates
coincide with the well known $S^{2}$ Bloch sphere coordinates :%

\begin{align}
X  &  =\left\langle \sigma_{x}\right\rangle _{\Psi}=2\operatorname{Re}%
(\overline{\alpha}\beta)\\
Y  &  =\left\langle \sigma_{y}\right\rangle _{\Psi}=2\operatorname{Im}%
(\overline{\alpha}\beta)\nonumber\\
Z  &  =\left\langle \sigma_{z}\right\rangle _{\Psi}=\left\vert \alpha
\right\vert ^{2}-\left\vert \beta\right\vert ^{2})\nonumber
\end{align}

This correspondance between Hopf map and Bloch sphere is not new
\cite{urtbanke}, but is poorly known in both communities (quantum optics and
geometry). It is striking that the simplest non trivial object of quantum
physics, the two-level system, bears such an intimate relation with the
simplest non trivial fibred space.

It is tempting to try to visualize the full ($S^{3}$) Hilbert space with its
fibre structure. This can be achieved by doing a (direct) stereographic map
from $S^{3}$ to $R^{3}$ (figure 1). Each $S^{3}$ circular fibre is mapped onto
a circle in $R^{3}$, with an exceptional straight line, image of the unique
$S^{3}$ great circle passing through the projection pole.

$%
{\parbox[b]{3.4523in}{\begin{center}
\includegraphics[
height=2.8055in,
width=3.4523in
]%
{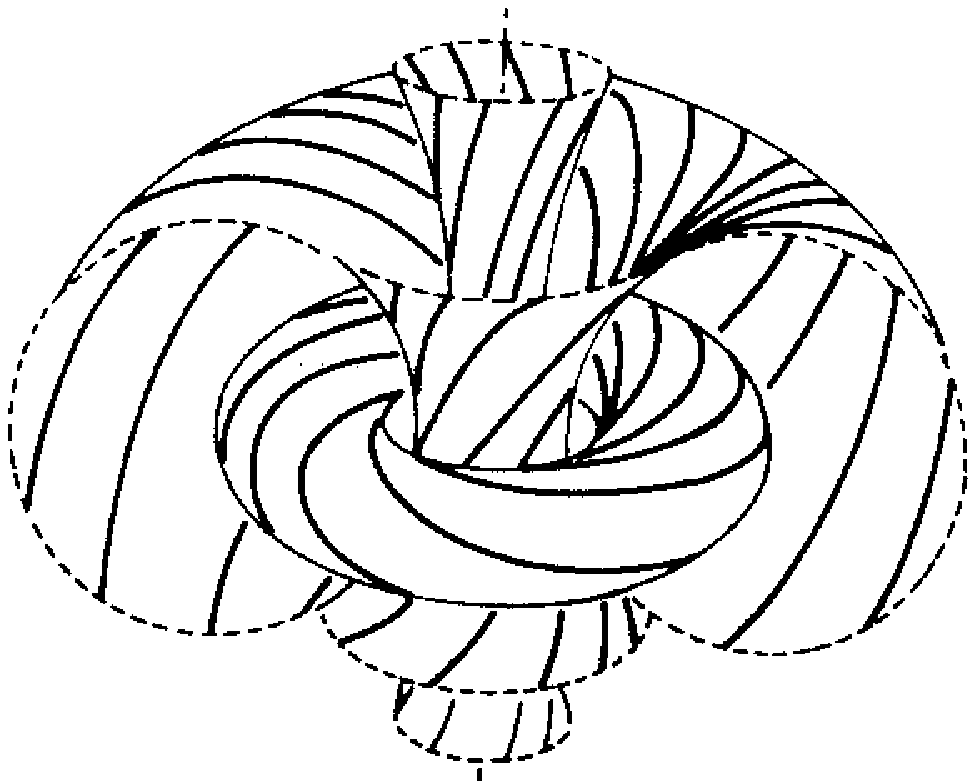}%
\\
Figure 1 : $S^3$ Hopf fibration after a stereographic map onto $R^3$. Circular
$S^1$ fibres are mapped onto circles in $R^3$, except the exceptional fibre
through the projection pole, which is mapped onto a vertical straight line.
Fibres can be grouped into a continuous family of nested tori, three of which
are shown here
\end{center}}}%
$

\qquad\qquad

\begin{center}

\end{center}

\section{ Two qubits, entanglement and the $S^{7}$ Hopf fibration}

\subsection{The two-qubits Hilbert space}

We now proceed one step further, and investigate pure states for two qubits.
The Hilbert space $\mathcal{E}$ for the compound system is the tensor product
of the individual Hilbert spaces $\mathcal{E}_{1}\otimes$ $\mathcal{E}_{2}$,
with a direct product basis $\left\{  \left\vert 00\right\rangle ,\left\vert
01\right\rangle ,\left\vert 10\right\rangle ,\left\vert 11\right\rangle
\right\}  $. A two-qubit state reads
\begin{align}
\left\vert \Psi\right\rangle  &  =\alpha\left\vert 00\right\rangle
+\beta\left\vert 01\right\rangle +\gamma\left\vert 10\right\rangle
+\delta\left\vert 11\right\rangle \qquad\label{two-qubits state}\\
\text{with \ }\alpha,\beta,\gamma,\delta &  \in\mathbb{C},\;\text{and
}\left\vert \alpha\right\vert ^{2}+\left\vert \beta\right\vert ^{2}+\left\vert
\gamma\right\vert ^{2}+\left\vert \delta\right\vert ^{2}=1\nonumber
\end{align}

$\left\vert \Psi\right\rangle $ is said \textquotedblright
separable\textquotedblright\ if it can be written as a simple product of
individual kets belonging to $\mathcal{E}_{1}$ and $\mathcal{E}_{2}$
separately, a definition which translates into the well known following
condition : $\alpha\delta=\beta\gamma$. A generic state is not separable, and
is said to be \textquotedblright entangled\textquotedblright. The $\left\vert
\Psi\right\rangle $ normalization condition $\left\vert \alpha\right\vert
^{2}+\left\vert \beta\right\vert ^{2}+\left\vert \gamma\right\vert
^{2}+\left\vert \delta\right\vert ^{2}=1$ identifies $\mathcal{E}$ to the
7-dimensional sphere $S^{7}$, embedded in $R^{8}$. It was therefore tempting
to see whether the known $S^{7}$ Hopf fibration (with fibres $S^{3}$ and base
$S^{4})$ can play any role in the Hilbert space description. This is the case
indeed, as we have shown recently \cite{mosseridandoloff}. Let us summarize
the main results, keeping in mind that some notations has been changed as
compared to this latter reference.

\subsection{The $S^{7}$ Hopf fibration}

One follows the same line as in the $S^{3}$ case, but using quaternions
instead of complex numbers (see appendix). We write
\begin{equation}
q_{1}=\alpha+\beta\mathbf{j},\;q_{2}=\gamma+\delta\mathbf{j},\qquad
q_{1,}q_{2}\in\mathbb{Q}\mathbf{,}%
\end{equation}

and a point (representing the state $\left\vert \Psi\right\rangle $) on the
unit radius $S^{7}$ as a pair of quaternions $\left(  q_{1,}q_{2}\right)  $
satisfying $\left\vert q_{1}\right\vert ^{2}+\left\vert q_{2}\right\vert
^{2}=1$. The Hopf map from $S^{7}$ to the base $S^{4}$ is the composition of a
map $h_{1}$ from $S^{7}$ to $R^{4}$ $(+\infty)$, followed by an inverse
stereographic map $h_{2}$ from $R^{4}$ to $S^{4}$.
\begin{align}
h_{1}  &  :%
\begin{array}
[c]{ccc}%
S^{7} & \longrightarrow & R^{4}+\left\{  \infty\right\} \\
\left(  q_{1},q_{2}\right)  \;\; & \longrightarrow & Q=\overline{q_{1}%
q_{2}^{-1}}%
\end{array}
\qquad q_{1,}q_{2}\in\mathbb{Q}\nonumber\\
h_{2}  &  :%
\begin{array}
[c]{ccc}%
R^{4}+\left\{  \infty\right\}  & \longrightarrow & S^{4}\\
Q & \longrightarrow & M(x_{l})
\end{array}
\qquad\sum\limits_{l=0}^{l=4}x_{l}^{2}=1
\end{align}

The base space $S^{4}$ is not embedded \ in $S^{7}$ : the fibration is again
not trivial. The fibre is a unit $S^{3}$ sphere as can seen easily by
remarking that the $S^{7}$ points $\left(  q_{1,}q_{2}\right)  $ and $\left(
q_{1}q,\,q_{2}q\right)  $, with $q$ a unit quaternion (geometrically a $S^{3}$
sphere) are mapped onto the same $Q$ value.

The $h_{1}$ map leads to%
\begin{align}
Q  &  =\overline{q_{1}q_{2}^{-1}}=\frac{1}{\sin^{2}\left(  \theta/2\right)
}\left[  \overline{\left(  \alpha+\beta\mathbf{j}\right)  \left(
\overline{\gamma}-\delta\mathbf{j}\right)  }\right]  =\frac{1}{\sin^{2}\left(
\theta/2\right)  }\left(  C_{1}+C_{2}\mathbf{j}\right) \\
\text{with }\sin\left(  \theta/2\right)   &  =\left\vert q_{2}\right\vert
\text{, }C_{1}=\left(  \overline{\alpha}\gamma+\overline{\beta}\delta\right)
,\text{ }C_{2}=\left(  \alpha\delta-\beta\gamma\right)  \text{ and }%
C_{1},C_{2}\in\mathbb{C}\nonumber
\end{align}

We face here a first striking result: \textbf{the Hopf map is entanglement
sensitive}! Indeed, non entangled states satisfy $\alpha\delta=\beta\gamma$
and therefore map onto the subset of pure complex numbers in the quaternion
field (both being completed by $\infty$ when the $Q$ denominator vanishes3).
Geometrically, this means that non-entangled states map from $S^{7}$ onto a
2-dimensional planar subspace of the target space $R^{4}$.

The second map $h_{2\text{ }}$sends states onto points on $S^{4}$, with
coordinates $x_{l}$, with $l$ running from to $0$ to $4$. With the inverse
stereographic pole located on the $S^{4}$ \textquotedblright north pole
\textquotedblright\ ($x_{0}=+1$), and the target space $R^{4\text{ }}$ cutting
$S^{4}$ along the equator, we get the following coordinate expressions%
\begin{align}
x_{0}  &  =\cos\theta=\left\vert q_{1}\right\vert ^{2}-\left\vert
q_{2}\right\vert ^{2}\label{S4_base}\\
x_{1}  &  =\sin\theta\;S(Q^{\prime})=2\operatorname{Re}\left(  \overline
{\alpha}\gamma+\overline{\beta}\delta\right) \nonumber\\
x_{2}  &  =\sin\theta\;V_{\mathbf{i}}(Q^{\prime})=2\operatorname{Im}\left(
\overline{\alpha}\gamma+\overline{\beta}\delta\right) \nonumber\\
x_{3}  &  =\sin\theta\;V_{\mathbf{j}}(Q^{\prime})=2\operatorname{Re}\left(
\alpha\delta-\beta\gamma\right) \nonumber\\
x_{4}  &  =\sin\theta\;V_{\mathbf{k}}(Q^{\prime})=2\operatorname{Im}\left(
\alpha\delta-\beta\gamma\right) \nonumber
\end{align}

$Q^{\prime}$ is the normalized image of the $h_{1}$ map ($Q^{\prime}%
=\tan\left(  \theta/2\right)  Q$), $S(Q^{\prime})$ and $V_{\mathbf{i}%
,\mathbf{j},\mathbf{k}}(Q^{\prime})$ being respectively the scalar and
vectorial parts of the quaternion $Q^{\prime}$ (see appendix). As for the
standard Bloch sphere case, the $x_{l}$ coordinates are also expectation
values of simple operators in the two-qubits state. An obvious one is $x_{0}$
which corresponds to $\left\langle \sigma_{z}\otimes Id\right\rangle _{\Psi}$.
The two next coordinates are also easily recovered as
\begin{align}
x_{1}  &  =2\operatorname{Re}\left(  \overline{\alpha}\gamma+\overline{\beta
}\delta\right)  =\left\langle \sigma_{x}\otimes Id\right\rangle _{\Psi}\\
x_{2}  &  =2\operatorname{Im}\left(  \overline{\alpha}\gamma+\overline{\beta
}\delta\right)  =\left\langle \sigma_{y}\otimes Id\right\rangle _{\Psi
}\nonumber
\end{align}

The remaining two coordinates, $x_{3}$ and $x_{4}$, are also expectation
values of an operator acting on $\mathcal{E}$, but in a more subtle way.
Define $\mathbf{J}$ as the (antilinear) \textquotedblright
conjugator\textquotedblright, an operator which takes the complex conjugate of
all complex numbers involved in an expression (here acting on the left in the
scalar product below). Form then the antilinear operator $\mathbf{E}$ (for
\textquotedblright entanglor\textquotedblright): $\mathbf{E}=-\mathbf{J}%
\left(  \sigma_{y}\otimes\sigma_{y}\right)  $. One finds%
\begin{align*}
x_{3}  &  =\operatorname{Re}\left\langle \mathbf{E}\right\rangle _{\Psi}\\
x_{4}  &  =\operatorname{Im}\left\langle \mathbf{E}\right\rangle _{\Psi}.
\end{align*}

Note that $\left\langle \mathbf{E}\right\rangle _{\Psi}$ vanishes for non
entangled states, and takes its maximal norm (equals to 1) for maximally
entangled states. \ Such an operator, which is nothing but the time reversal
operator for two spins
${\frac12}$%
, is already widely used in quantifying entanglement\cite{wootters}, through a
quantity called the \textquotedblright concurrence\textquotedblright\ $c$,
which corresponds here to $c=2\left\vert C_{2}\right\vert $.

\subsection{Generalized Bloch sphere for the two-qubit case}

Let us first inverse the Hopf map, and get the general expression for the set
of states ( a $S^{3}$ sphere in $S^{7}$) which is sent to $Q$ by the $h_{1}$
map. A generic such state, noted $\Psi_{Q}$, reads (given as a pair of quaternions)%

\begin{equation}
\Psi_{Q}=(\cos\left(  \theta/2\right)  \,q\,,\sin\left(  \theta/2\right)
Q^{\prime}\,q), \label{Psi_Q1}%
\end{equation}

with $q$ a unit quaternion spanning the $S^{3}$ fibre. Notice that we could
also write $\Psi_{Q}$ in a way that recall the standard spinor notation (but
here with quaternionic instead of complex components):
\begin{equation}
\Psi_{Q}=(\cos\left(  \theta/2\right)  \exp\left(  -\varphi\mathbf{t}%
/2\right)  \,q\,,\sin\left(  \theta/2\right)  \exp\left(  \varphi
\mathbf{t}/2\right)  \,q), \label{Psi_Q2}%
\end{equation}

where $\cos\varphi=x_{1}/\sin\theta=S(Q^{\prime})$, and $\mathbf{t}$ is the
following unit pure imaginary quaternion :%
\begin{equation}
\mathbf{t=}\left(  \mathbf{V}_{\mathbf{i}}(Q^{\prime})\mathbf{i}%
+\mathbf{V}_{\mathbf{j}}(Q^{\prime})\mathbf{j}+\mathbf{V}_{\mathbf{k}%
}(Q^{\prime})\mathbf{k}\right)  /\sin\varphi. \label{unit_pure}%
\end{equation}

In order to compare with the generic expression
({\normalsize \ref{two-qubits state}}), we aim to write $\Psi_{Q}$ as a
quadruplet of complex numbers. For that purpose, we express the two unit
quaternions $q$ and $Q^{\prime}$ in terms of pairs of complex numbers,
$q=a+b\mathbf{j}$ (with $\left\vert a\right\vert ^{2}+\left\vert b\right\vert
^{2}=1$), and $Q^{\prime}=u+v\mathbf{j}$ (with $\left\vert u\right\vert
^{2}+\left\vert v\right\vert ^{2}=1$), and eventually get:%
\begin{equation}
\Psi_{Q}=(\cos\left(  \theta/2\right)  a\,,\cos\left(  \theta/2\right)
b,\sin\left(  \theta/2\right)  (ua-v\overline{b}),\sin\left(  \theta/2\right)
(ub+v\overline{a})) \label{Psi_Q3}%
\end{equation}

In the above expression, $\theta,$ $u$ and $v$ correspond to the base space
part of the fibration. Furthermore, we can relate $u$ and $v$ to already known
quantities. Indeed%
\[
u=(x_{1}+\mathbf{i}x_{2})/\sin\theta=\left\langle \left(  \sigma
_{x}+\mathbf{i}\sigma_{y}\right)  \otimes Id\right\rangle _{\Psi}/\sin\theta.
\]

In addition, the state global phase indeterminacy allows to take $v$ a real.
More precisely%
\[
v=c/\sin\theta\text{\ , where }c\text{ is the above mentionned concurrence}%
\]

Let us now describe the two extreme cases of separable and maximally entangled states.

\subsubsection{Separable states}

In the non entangled case, we have seen above that $Q$ is a complex number,
$\mathbf{t}=\mathbf{i,}$ and therefore $u=\exp\mathbf{i}\varphi$ and $v=0$.
The above expression simplifies to%
\begin{equation}
\Psi_{Q}=(\cos\left(  \theta/2\right)  \,a,\cos\left(  \theta/2\right)
b,\sin\left(  \theta/2\right)  \,a\exp\mathbf{i}\varphi,\sin\left(
\theta/2\right)  b\exp\mathbf{i}\varphi). \label{Psi_sep1}%
\end{equation}

Up to a global rescaling by $\exp\left(  -\mathbf{i}\varphi/2\right)  $, one
gets the following ket $\left\vert \Psi_{Q}\right\rangle $:%
\begin{equation}
\left\vert \Psi_{Q}\right\rangle =\left(  \cos\left(  \theta/2\right)
\exp\left(  -\mathbf{i}\varphi/2\right)  \left\vert 0\right\rangle _{1}%
+\sin\left(  \theta/2\right)  \exp\left(  \mathbf{i}\varphi/2\right)
\left\vert 1\right\rangle _{1}\right)  \otimes\left(  \,a\left\vert
0\right\rangle _{2}+b\left\vert 1\right\rangle _{2}\right)  \label{Psi_sep2}%
\end{equation}

The projective Hilbert space for two non-entangled qubits is known to be the
product of two 2-dimensional spheres $S_{1}^{2}\times S_{2}^{2}$, each sphere
being the Bloch sphere associated with the given qubit. This property is
clearly displayed here. The unit $S^{4}$ base space reduces to a unit $S^{2}$
sphere (since $x_{3}=x_{4}=0)$ which is nothing but the Bloch sphere for the
first qubit. The second qubit Bloch sphere is then recovered from the fibre,
spanned by $q=a+b\mathbf{j}$. \ Indeed, we can iterate the fibration process
on the $S^{3}$ fibre itself and get the (Hopf fibration base)-(Bloch sphere)
coordinates for this two-level system.\ It is now easy to recover that this
new $S^{2}$ base is the second qubit Bloch sphere.

In summary, for non entangled qubits, the $S^{7}$ Hopf fibration, with base
$S^{4}$ and fibre $S^{3}$, simplifies to the simple product of a $S^{2}$
sub-sphere of the base (the first qubit Bloch sphere) by a second $S^{2}$ (the
second qubit Bloch sphere) obtained as the base of a $S^{3}$ Hopf fibration
applied to the fibre itself. Let us stress that this last iterated fibration
is necessary to take into account the global phase of the two qubit system.

The fact that these two $S^{2}$ spheres play a symmetrical role (although one
is related to the base and the other to the fibre) can be understood in the
following way. We grouped together $\alpha$ and $\beta$ on one hand, and
$\gamma$ and $\delta$ on the other hand, to form the quaternions $q_{1}$ and
$q_{2}$, and then define the Hopf map $h_{1}$ as the ratio of these two
quaternions (plus a complex conjugation). Had we grouped $\alpha$ and $\gamma
$, and $\beta$ and $\delta$, to form two new quaternions, and use the same
definition for the Hopf map, we would also get a $S^{7}$ Hopf fibration, but
differently oriented. We let as an exercise to compute the base and fibre
coordinates in that case. The net effect is to interchange the role of the two
qubits: the second qubit Bloch sphere is now part of the $S^{4}$ base, while
the first qubit Bloch sphere is obtained from the $S^{3}$ fibre.

\subsubsection{Maximally entangled states}

Let us now focus on maximally entangled states (M.E.S.). They correspond to
the complex number $C_{2}$ having maximal norm $1/2$ (unit concurrence). This
in turn implies that the Hopf map base reduces to a unit circle in the plane
$\left(  x_{3},x_{4}\right)  $, parametrized by the unit complex number
$2C_{2}$. The projective Hilbert space for these M.E.S. is known to be
$S^{3}/Z_{2},$ a $S^{3}$ \ sphere with identified opposite points
\cite{2qbits-geometry} (this is linked to the fact that all M.E.S. can be
related by a local operation on one sub-system, since $S^{3}/Z_{2}=SO(3)$). In
order to recover this result in the present framework, one can follow the
trajectory of a representative point on the base and on the fibre while the
state is multiplied by an overall phase $\exp\left(  i\omega\right)  $. The
expression for $C_{2}\left(  =\alpha\delta-\beta\gamma\right)  $ shows that
the point on the base turns by twice the angle $\omega$. Only when $\omega
=\pi$ does the corresponding state belongs to the same fiber (e.g. maps onto
the same value on the base). The fact that the fibre is a $S^{3}$ sphere, and
this two-to-one correspondance between the fibre and the base under a global
phase change, explains the $S^{3}/Z_{2}$ topology for the M.E.S. projective
Hilbert space. Let us give now a more explicit proof of that result.

M.E.S.correspond to $\theta=\pi/2$ and maximal concurrence ($c=1$), which
leads to $u=0$ and $v=1$. $\Psi_{MES}$ therefore read , from expression
($\ref{Psi_Q3}$):
\begin{equation}
\Psi_{MES}=\frac{1}{\sqrt{2}}(a,b\,,-\overline{b},\overline{a}).
\label{Psi_MES1}%
\end{equation}

The latter expression ($\ref{Psi_MES1}$) for maximally entangled states is
rather interesting in that it directly shows the $S^{3}/Z_{2}$ topology for
the M.E.S. projective Hilbert space. Indeed, the M.E.S$.$ set corresponds to
pairs $\left(  a,b\right)  $, which as a whole cover a unit radius $S^{3}$
sphere. Now, looking to the quadruplet expression ($\ref{Psi_MES1}),$ opposite
points $(a,b)$ and $(-a,-b)$ on $S^{3}$ clearly correspond to the same state
$\Psi_{MES}$ (up to a global phase). Opposite points on $S^{3}$ have therefore
tobe identified, leading to the $S^{3}/Z_{2}$ ($\equiv SO(3)$) structure.

This one-to-one correspondance between M.E.S. and 3 dimensional rotation
matrices has recently led to propose using the former in an "applied topology"
experiment \cite{milman-mosseri03}: to verify experimentally the well known
subtle topology of the (two-fold connected) $SO(3)$ group. The latter property
is evidenced by constructing the two inequivalent family of closed paths in
the geometrical manifold representing this group. This is done by choosing
sequences of unitary operations on the MES two-qubits states. The non
equivalence between the two paths is manifested by a $\pi$ topological phase
shift which should result from an adequate interference experiment (a twin
photons experiment have been proposed, but other two-qubit states could be used).

\subsubsection{A generalization of the Bloch sphere representation}

We are now led to consider a generalization of the Bloch sphere for the
two-qubit projective Hilbert space. Clearly, the present Hopf fibration
description suggests a splitting of the representation space in a product of
base and fibres sub-spaces. Of the base space $S^{4}$, we propose to only keep
the first three coordinates
\begin{equation}
\left(  x_{0},x_{1},x_{2}\right)  =\left(  \left\langle \sigma_{z}\otimes
Id\right\rangle _{\Psi},\left\langle \sigma_{x}\otimes Id\right\rangle _{\Psi
},\left\langle \sigma_{y}\otimes Id\right\rangle _{\Psi}\right)
\end{equation}

All states map inside a standard ball $B^{3}$ of radius 1, where the set of
separable states forms the $S^{2}$ boundary (the usual first qubit Bloch
sphere), and the centre corresponds to maximally entangled states. Concentric
spherical shells around the centre correspond to states of equal concurrence
$c$ (maximal at the centre, zero on the surface), the radius of the spherical
shell being equal to $\sqrt{1-c^{2}}$. The idea of slicing the 2-qubit Hilbert
space into manifolds of equal concurrence is not new \cite{2qbits-geometry}%
,\cite{CPfoliation}. What is nice here is that, under the Hopf map (and a
projection onto the 3d subspace of the base spanned by the first three
coordinates), these manifolds transforms into concentric $S^{2}$ shells which
fill the unit ball.

To each point ($x_{0},x_{1},x_{2}$), it corresponds a $S^{3}/Z_{2}$ manifold,
spanned by the couple ($a,b$), as seen clearly from relation ($\ref{Psi_Q3}$),
with an added identification of ($a,b$) and ($-a,-b$). The natural
generalization of the Bloch sphere for two qubits is therefore a product of
two $B^{3}$ balls. The first one, spanned by the triple ($x_{0},x_{1},x_{2}$),
has just been described as containing the partial Bloch sphere for one the two
qubits, with its set of concentric iso-concurrence spheres. The second one
corresponds to the standard representation of $SO(3)$ by a
$\overleftrightarrow{B^{3}}$ ball of radius $\pi$, the double arrow sign
recalling that opposite points on the boundary $S^{2}$ sphere have to be
identified. This picture is valid for all states except the separable ones,
for which the fiber derived $\overleftrightarrow{B^{3}}$ space reduces to a
$S^{2}$ sphere (the second qubit partial Bloch sphere).

Instead of mapping the continuous set of $S^{2}$ spheres onto the filled ball
$B^{3}$ in the space spanned by ($x_{0},x_{1},x_{2}$), this nice(but singular)
foliation of the two-qubit projective Hilbert space (here the complex
projective space $CP^{3})$ can be also pictured as a concurrence segment
(between 0 and 1) with the corresponding sub-manifolds. The sub-space of
vanishing concurrence has a $S^{2}\times S^{2}$ structure, while that of
maximal ($c=1$) concurrence corresponds to $SO(3)$. Sub-manifolds of
intermediate concurrence have the structure of a direct product $S^{2}\times
SO(3)$, the sphere $S^{2}$ having radius $\sqrt{1-c^{2}}$. This illustrated in
the figure 2 below

\begin{center}%
\begin{figure}
[ptb]
\begin{center}
\includegraphics[
height=1.2341in,
width=3.435in
]%
{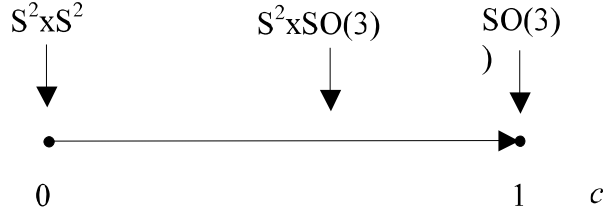}%
\caption{Foliation of the two-qubits Hilbert space with respect to state
entanglement}%
\end{center}
\end{figure}
\bigskip\bigskip
\end{center}

\textbf{Examples}\linebreak

Back to to the $B^{3}$ picture, let us for instance go through the states
along a simple $B^{3}$ ray, from the point $\left(  0,1,0\right)  ,$ image of
the states $\Psi_{S}$ such that $\left\langle \sigma_{x}\otimes
Id\right\rangle _{\Psi}=1$, to the $B^{3}$ center $\left(  0,0,0\right)  $,
image the maximally entangled states $\Psi_{MES}$ . $Q^{\prime}$ reduces to a
unit circle in the plane spanned by ($x_{1},x_{3}$), and these states are
therefore parametrized by a single angle $\epsilon$ in the interval $\left[
0,\pi/2\right]  $,%

\begin{equation}
\Psi_{\epsilon}=\frac{1}{\sqrt{2}}(\,q\,,\exp\left(  \epsilon\mathbf{j}%
\right)  \,\,q).\text{,} \label{Psi_eps1}%
\end{equation}

or, written as a quadruplet of complex numbers,%

\begin{equation}
\Psi_{\epsilon}=\frac{1}{\sqrt{2}}(a\,,b,a\cos\epsilon-\overline{b}%
\sin\epsilon,b\cos\epsilon+\overline{a}\sin\epsilon). \label{Psi_eps2}%
\end{equation}

For $\epsilon=0$, one gets%
\begin{align}
\Psi_{\epsilon=0}  &  =\Psi_{S}=\frac{1}{\sqrt{2}}\left(  a,b,a,b\right)
.\text{ and therefore}\label{Psi_eps3}\\
\left\vert \Psi_{S}\right\rangle  &  =\frac{1}{\sqrt{2}}\left(  \left\vert
0\right\rangle _{1}+\left\vert 1\right\rangle _{1}\right)  \otimes\left(
a\left\vert 0\right\rangle _{2}+b\left\vert 1\right\rangle _{2}\right)
,\nonumber
\end{align}

as expected for the set of separable states which are eigenstates of
$\sigma_{x}\otimes Id$ (with eigenvalue $+1$).

For $\epsilon=\pi/2$, the above set of maximally entangled states, as given by
relation ($\ref{Psi_MES1}$), is recovered. Intermediate values of $\epsilon$
correspond to less entangled states, whose concurrence read $c=\sin\epsilon$,
as can be easily found from equation ($\ref{Psi_eps2}$). Expression
($\ref{Psi_eps2}$) also proves that the set of such states describes a
$S^{3}/Z_{2}$ manifold.

Similar analyses can be done for any path inside the $B^{3}$ ball. A second
very simple example is provided by the path from $\left(  1,0,0\right)  $ to
$\left(  0,0,0\right)  $. In that case the states, again parametrized by an
angle $\epsilon$, read%
\[
\Psi_{\epsilon}=(\cos\frac{\epsilon}{2}a\,,\cos\frac{\epsilon}{2}b,-\sin
\frac{\epsilon}{2}\overline{b},\sin\frac{\epsilon}{2}\overline{a}).
\]

\subsubsection{Relation with the Bloch ball representation for mixed states}

The Bloch ball single qubit mixed state representation was recalled above. In
that case, the centre of the Bloch ball corresponds to maximally mixed states.
The reader should not be surprised to find here (in the two-qubits case) a
second unit radius ball, with maximally entangled states now at the centre. It
corresponds to a known relation between partially traced two-qubit pure states
and one-qubit mixed state. Indeed the partially traced density matrix
$\rho_{1}$ is simply written in terms of $C_{1}$ and $C_{2}$ derived from the
$S^{7}$ Hopf map:%

\begin{equation}
\rho_{1}=\frac{1}{2}\left(
\begin{array}
[c]{cc}%
1+x_{0} & x_{1}-\mathbf{i\,}x_{2}\\
x_{1}+\mathbf{i\,}x_{2} & 1-x_{0}%
\end{array}
\right)  =\left(
\begin{array}
[c]{cc}%
\left\vert q_{1}\right\vert ^{2} & \overline{C_{1}}\\
C_{1} & \left\vert q_{2}\right\vert ^{2}%
\end{array}
\right)  \text{ \ }%
\end{equation}

with unit trace and $\det\rho_{1}=\left\vert C_{2}\right\vert ^{2}.$ The
partial $\rho_{1}$ represents a pure state density matrix whenever $C_{2}$
vanishes (the separable case), and allows for a unit Bloch sphere (that
associated to the first qubit). It corresponds to a mixed state density matrix
as soon as $\left\vert C_{2}\right\vert >0$ (and an entangled state for the
two qubit state). The other partially traced density matrix $\rho_{2}$ is
related to the other $S^{7}$ Hopf fibration which was discussed above.

\section{ Three qubits, and the $S^{15}$ Hopf fibration}

\subsection{Three qubits}

The Hilbert space $\mathcal{E}$ for the compound system is the tensor product
of the individual Hilbert spaces $\mathcal{E}_{1}\otimes$ $\mathcal{E}%
_{2}\otimes$ $\mathcal{E}_{3}$, with a direct product basis
\[
\left\{  \left\vert 000\right\rangle ,\left\vert 001\right\rangle ,\left\vert
010\right\rangle ,\left\vert 011\right\rangle ,\left\vert 100\right\rangle
,\left\vert 101\right\rangle ,\left\vert 110\right\rangle ,\left\vert
111\right\rangle \right\}  ,
\]
which can be written $\left\{  \left\vert l\right\rangle ,l=0..7\right\}  $. A
three-qubit state reads
\begin{equation}
\left\vert \Psi\right\rangle =\sum_{l=0}^{7}t_{l}\left\vert l\right\rangle
\text{\ \ \ with \ }t_{l}\in\mathbb{C},\;\text{and }\sum\text{ }\left\vert
t_{l}\right\vert ^{2}=1\nonumber
\end{equation}

The $\left\vert \Psi\right\rangle $ normalization condition identifies
$\mathcal{E}$ to the 15-dimensional sphere $S^{15}$, embedded in $R^{16}$.
This suggests looking to how far the third Hopf fibration (that of $S^{15}$,
with base $S^{8}$ and fibres $S^{7})$ can be helpful for describing the 3
qubits Hilbert space geometry.

\subsection{The $S^{15}$ Hopf fibration}

One proceeds along the same line as for the previous $S^{3}$ and $S^{7}$
cases, but using now octonions (see appendix). We write
\begin{equation}
a=a^{\prime}+a"\mathbf{e},\;b=b^{\prime}+b"\mathbf{e},\qquad a,b\in
\mathbb{O}_{,}\text{ and }a^{\prime},a",b^{\prime},b"\in\mathbb{Q}\mathbf{,}%
\end{equation}

and a point (representing the state $\left\vert \Psi\right\rangle $) on the
unit radius $S^{15}$ as a pair of octonions $\left(  a,b\right)  $ satisfying
$\left\vert a\right\vert ^{2}+\left\vert b\right\vert ^{2}=1$. But, to get a
Hopf map of physical interest, with coordinates simply related to interesting
observable expectation values, one needs to define a slightly tricky relation
between $\left\vert \Psi\right\rangle $ and the octonions pair $\left(
a,b\right)  $, as follows:%
\begin{align}
a  &  =(t_{0}+t_{1}\mathbf{j,}t_{2}+\mathbf{j}t_{3})=(t_{0}+t_{1}%
\mathbf{j,}t_{2}+\overline{t_{3}}\mathbf{j})=\left(  a^{\prime},a^{\prime
\prime}\right) \label{3qubit_octo}\\
b  &  =(t_{4}+t_{5}\mathbf{j,}t_{6}+\mathbf{j}t_{7})=(t_{4}+t_{5}%
\mathbf{j,}t_{6}+\overline{t_{7}}\mathbf{j})=\left(  b^{\prime},b^{\prime
\prime}\right) \nonumber
\end{align}

The Hopf map from $S^{15}$ to the base $S^{8}$ is the composition of a map
$h_{1}$ from $S^{15}$ to $R^{8}$ $(+\infty)$, followed by an inverse
stereographic map $h_{2}$ from $R^{8}$ to $S^{8}$.
\begin{align}
h_{1}  &  :%
\begin{array}
[c]{ccc}%
S^{15} & \longrightarrow & R^{8}+\left\{  \infty\right\} \\
\left(  a,b\right)  \;\; & \longrightarrow & P=\overline{ab^{-1}}%
\end{array}
\qquad a,b\in\mathbb{O}\nonumber\\
h_{2}  &  :%
\begin{array}
[c]{ccc}%
R^{8}+\left\{  \infty\right\}  & \longrightarrow & S^{8}\\
P & \longrightarrow & M(x_{l})
\end{array}
\qquad\sum\limits_{l=0}^{l=8}x_{l}^{2}=1
\end{align}

The base space $S^{8}$ is not embedded \ in $S^{15}$ : the fibration is again
not trivial.

The fibre is a unit $S^{7}$ sphere, the proof of which is more tricky (and not
given here) than in the lower dimension case. The $h_{1}$ map leads to%
\begin{align}
P  &  =\overline{ab^{-1}}=\frac{1}{\sin^{2}\theta/2}\left(  Q_{1}%
+Q_{2}\mathbf{e}\right) \\
\text{with }\sin\theta/2  &  =\left\vert b\right\vert \text{, }Q_{1}%
=(b^{\prime}\overline{a^{\prime}}+\overline{a^{\prime\prime}}b^{\prime\prime
}),\text{ }Q_{2}=\left(  -a^{\prime\prime}b^{\prime}+b^{\prime\prime}%
a^{\prime}\right)  \text{ and }Q_{1},Q_{2}\in\mathbb{Q}\nonumber
\end{align}

Athough this is not at first sight evident, the Hopf map is still entanglement
sensitive in that case. To show this, it is instructive to first express
$Q_{1}$ and $Q_{2}$ in term of the $t_{l}$ components read out from
($\ref{3qubit_octo}).$%

\begin{align*}
Q_{1}  &  =\left(  \overline{t_{0}}t_{4}+\overline{t_{1}}t_{5}+\overline
{t_{2}}t_{6}+\overline{t_{3}}t_{7})+(t_{0}t_{5}-t_{1}t_{4}+\overline{t_{2}%
}\overline{t_{7}}-\overline{t_{3}}\overline{t_{6}}\right)  \,\mathbf{j}\\
Q_{2}  &  =\left(  t_{0}t_{6}+t_{2}t_{4}+\overline{t_{3}}\overline{t_{5}%
}-\overline{t_{1}}\overline{t_{7}})+(t_{1}t_{6}-t_{2}t_{5}+\overline{t_{0}%
}\overline{t_{7}}-\overline{t_{3}}\overline{t_{4}}\right)  \,\mathbf{j}%
\end{align*}

Let us introduce the generalised complex concurrence terms $T_{ij,kl}%
=t_{i}t_{j}-t_{k}t_{l}$. They allow to write in a synthetic form the
coordinates on the base $S^{8}.$ The second map $h_{2\text{ }}$sends states
onto points on $S^{8}$, with coordinates $x_{l}$, with $l$ running from to $0$
to $8$. With the inverse stereographic pole located on the $S^{8}$
\textquotedblright north pole \textquotedblright\ ($x_{0}=+1$), and the target
space $R^{8\text{ }}$ cutting $S^{8}$ along the equator, we get the following
coordinate expressions%
\begin{align}
x_{0}  &  =\cos\theta=\left\vert a\right\vert ^{2}-\left\vert b\right\vert
^{2}=\left\langle \sigma_{z}\otimes Id\otimes Id\right\rangle _{\Psi}\\
x_{1}+\mathbf{i\,}x_{2}  &  =2\left(  \overline{t_{0}}t_{4}+\overline{t_{1}%
}t_{5}+\overline{t_{2}}t_{6}+\overline{t_{3}}t_{7}\right)  =\left\langle
\left(  \sigma_{x}+\mathbf{i\,}\sigma_{y}\right)  _{1}\otimes Id\otimes
Id\right\rangle _{\Psi}\nonumber\\
x_{3}+\mathbf{i\,}x_{4}  &  =2\left(  T_{05,14}+\overline{T_{27,36}}\right)
\nonumber\\
x_{5}+\mathbf{i\,}x_{6}  &  =2\left(  T_{06,24}+\overline{T_{35,17}}\right)
\nonumber\\
x_{7}+\mathbf{i\,}x_{8}  &  =2\left(  T_{16,25}+\overline{T_{07,34}}\right)
\end{align}

A lengthy, but trivial, computation allows to verify that \ the base $S^{8}$
has unit radius.

\subsection{Discussion}

It is easy to show that three-qubits states such that the first qubit is
separated from the two others map onto a point such that $x_{j}=0,$ for
$j=4,5,6,7,8.$ One way to show this is to realize that, in a multi-qubit
state, a given qubit is separated from the others when its partial Bloch
sphere has a unit radius. The first qubit partial Bloch sphere is spanned here
by the triplet $(x_{0},x_{1},\mathbf{\,}x_{2}).$ A second proof consists in
writing down the separability algebraic conditions \cite{jorrand}. In the
present case, the latter imply that the above generalized concurrences vanish
(in fact the six vanishing conditions only rely onto three independant
conditions). Going back to the above definition of the $h_{1}$ map, this means
that in that case, the Hopf map carries an octonion couple onto a pure complex
number $P$. Therefore, as for two-qubits and $S^{7}$, \textbf{the
}$S^{15\text{ }}$\textbf{Hopf fibration is also entanglement sensitive for
three qubits}! This result has been independantly derived by Bernevig and
Chen\cite{bernevig}.

However, one should notice an important difference between the two- and
three-qubit cases. In the two-qubit case, the $S^{7}$ Hopf fibration have
allowed us to foliate the projective Hilbert space with respect to state
entanglement, the latter, measured by the concurrence, being simply related to
the norm the restriction of the base point to the subspace spanned by the
triplet $(x_{3},x_{4}).$ Since the base space $S^{4}$ has unit radius, the
entanglement is therefore simply related to the radius the first qubit partial
Bloch sphere, spanned by the triplet $(x_{0},x_{1},\mathbf{\,}x_{2})$. The
first-to-second qubit distinction (base-fibre in the fibration) do not matter
here (to define the foliation) since the two partial Bloch sphere radii are equal.

The $S^{15\text{ }}$Hopf fibration is clearly sensitive to the entanglement of
one qubit (the first qubit in the present case) with respect to the other two
qubits: the first qubit partial Bloch sphere radius is still read out from the
norm of the restriction of the base point to the subspace spanned by the
triplet $(x_{0},x_{1},\mathbf{\,}x_{2})$. However, the latter does not tell
the whole story in terms of state entanglement. Once you know how far qubit 1
is entangled with the remaining two, you do not yet know whether the second
(or third) qubit is or not separated. This prevents from building a foliation
driven by a single entanglement parameter. One possibility for such a
foliation unique parameter would be to use the 3-tangle\cite{3-tangle}. But it
does not distinguish among separable and entangled W states. An alternative
entanglement parameter has been by suggested by Bernevig and Chen
\cite{bernevig} (see also ref.\cite{meyer2002} for a related measure): to
recover the symmetry between the three qubits, an average over the three
qubits partial Bloch sphere radii is used.

The solution to the foliation problem might be to use three (instead of one)
parameters, by considering three distinct $S^{15\text{ }}$Hopf fibrations,
such that each of the three qubits partial Bloch spheres is singled out by the
first three coordinates on the base. Said more simply, one may try to describe
the Hilbert space geometry in a space spanned by the three partial Bloch
sphere radii ($r_{1},r_{2},r_{3}$). Since each radius belongs to the interval
$\left[  0,1\right]  ,$ the whole representation lies inside a unit cube. When
one radius equals 1, one qubit is separated from the other two, and the other
two radii are equal. We then see that the set under consideration cuts three
square faces of the cube (such that $r_{j}=1),$ along a diagonal. These
diagonals corresponds to the product of a sphere $S^{2}$ (the Bloch sphere of
the separated qubit), and a copy of the above foliation for the remaining two
qubits. Notice that the latter foliation was introduced above with respect to
state entanglement, as measured by the concurrence $c$, instead of the partial
Bloch sphere radius. To get a fully equivalent picture, one should therefore
transform the original coordinate system from $c$ to $r=\sqrt{1-c^{2}}$.

\section{Conclusion}

Our main goal in this paper was to provide a geometrical representation of the
two and three-qubit Hilbert space pure states. We have shown that, as for the
one-qubit Bloch sphere relation to the $S^{3}$ Hopf fibration, the more
complex $S^{7}$ and $S^{15}$ Hopf fibrations also play a natural role in that
case. Note that, as already mentionned in ref. \cite{mosseridandoloff}, in the
three Hopf fibration sequence ($S^{15},S^{7},S^{3})$, the fibre in the larger
dimensional space is the full space in next case. This is illustrated below:

\bigskip%
\begin{figure}
[ptb]
\begin{center}
\includegraphics[
height=2.2684in,
width=4.7902in
]%
{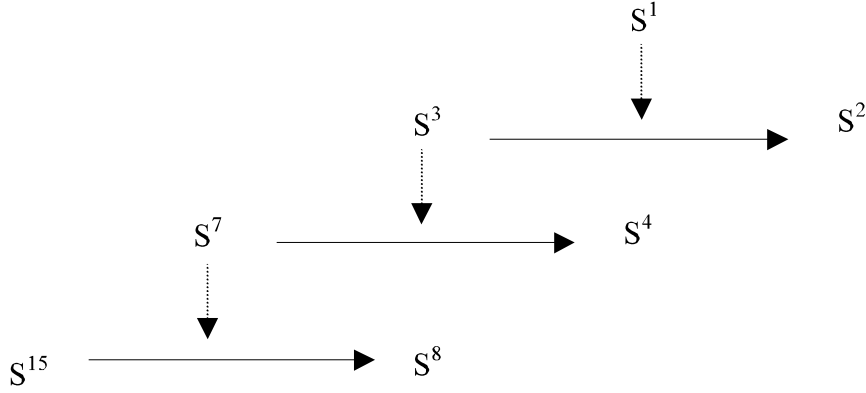}%
\caption{Figure 3: Nesting of the three Hopf fibrations}%
\end{center}
\end{figure}

This offers the possibility of further nesting the fibrations, a possibility
that was already used in the present analysis of two-qubit separable case.
This also applies to the three qubit case: whenever the first qubit is
separated from the other two, the base reduces to the first qubit Bloch
sphere, and the $S^{7}$ fibre is precisely the Hilbert space of the remaining
two qubits.

In the two-qubit case, this approach has in particular allowed for a complete
description of the pure state projective Hilbert space, in terms of an
entanglement driven foliation, with well characterized leaves. This goal has
not yet been achieved in the three qubits case, although a plausible track has
been proposed here.%

\appendix

\section{\textbf{Quaternions and Octonions}}

\begin{center}
\textbf{Quaternions}
\end{center}

Quaternions are usually presented with the imaginary units $\mathbf{i}%
,\mathbf{j}$ et $\mathbf{k}$ in the form~:%
\[
q=x_{0}+x_{1}\mathbf{i}+x_{2}\mathbf{j}+x_{3}\mathbf{k},\qquad x_{0}%
,x_{1},x_{2},x_{3}\in\mathbb{R}\quad\text{with }\mathbf{i}^{2}=\mathbf{j}%
^{2}=\mathbf{k}^{2}=\mathbf{ijk}=-1,
\]

the latter \textquotedblright Hamilton\textquotedblright\ relations defining
the quaternion multiplication rules which are non-commutative. They can also
be defined equivalently, using the complex numbers $c_{1}=x_{0}+x_{1}%
\mathbf{i}$ and $c_{2}=x_{2}+x_{3}\mathbf{i}$, in the form $q=c_{1}%
+c_{2}\mathbf{j}$, or equivalently as an ordered pair of complex numbers
staisfying
\begin{align*}
\left(  c_{1},c_{2}\right)  +\left(  d_{1},d_{2}\right)   &  =\left(
c_{1}+d_{1},c_{2}+d_{2}\right) \\
\left(  c_{1},c_{2}\right)  \left(  d_{1},d_{2}\right)   &  =\left(
c_{1}d_{1}-c_{2}\overline{d_{2}},c_{1}d_{2}+c_{2}\overline{d_{1}}\right)
\end{align*}

The conjugate of a quaternion $q$ is $\overline{q}=x_{0}-x_{1}\mathbf{i}%
-x_{2}\mathbf{j}-x_{3}\mathbf{k=}\overline{c_{1}}-c_{2}\mathbf{j}$ and its
squared norm reads $N_{q}^{2}=q\overline{q}$.

Another way in which $q$ can be written is as a scalar part $S(q)$ and a
vectorial part $\mathbf{V}(q)$:%
\[
q=S(q)+\mathbf{V}(q),\;S(q)=x_{0},\;\mathbf{V}(q)=x_{1}\mathbf{i}%
+x_{2}\mathbf{j}+x_{3}\mathbf{k,}%
\]

with the relations%
\[
S(q)=\frac{1}{2}(q+\overline{q}),\;\mathbf{V}(q)=\frac{1}{2}(q-\overline{q}).
\]

A quaternion is said to be real if $\mathbf{V}(q)=0$, and pure imaginary if
$S(q)=0$. We shall also write$\;\mathbf{V}_{\mathbf{i},\mathbf{j},\mathbf{k}%
}(q)$ for the component of $\mathbf{V}(q)$ along $\mathbf{i},\mathbf{j}%
,\mathbf{k.}$ Finally, and as for complex numbers, a quaternion can be noted
in an exponential form as%
\[
q=\left|  q\right|  \exp\varphi\mathbf{t=}\left|  q\right|  \left(
\cos\varphi+\sin\varphi\quad\mathbf{t}\right)  \text{, }%
\]

where $\mathbf{t}$ is a unit pure imaginary quaternion. When $\mathbf{t}%
=\mathbf{i}$, usual complex numbers are recovered. Note that quaternion
multiplication is non-commutative so that%
\[
\exp\varphi\mathbf{t}\exp\lambda\mathbf{u=\exp(}\varphi\mathbf{t+}%
\lambda\mathbf{u)}%
\]

only if $\mathbf{t=u}$.

\begin{center}
\textbf{Octonions}
\end{center}

An octonion $a$ can be defined by introducing a new unit $\mathbf{e}$
(different from the preceding unit quaternions $\mathbf{i},\mathbf{j}$ and
$\mathbf{k}$, and such that $\mathbf{e}^{2}=-1)$, and pairs of quaternions
$a^{\prime},a^{\prime\prime}$:%

\begin{align*}
a  &  =a^{\prime}+a^{\prime\prime}\mathbf{e}\\
ab  &  =\left(  a^{\prime}+a^{\prime\prime}\mathbf{e}\right)  \left(
b^{\prime}+b^{\prime\prime}\mathbf{e}\right) \\
&  =\left(  a^{\prime}b^{\prime}-\overline{b^{\prime\prime}}a^{\prime\prime
}\right)  +\left(  b^{\prime\prime}a^{\prime}+a^{\prime\prime}\overline
{b^{\prime}}\right)  \mathbf{e}%
\end{align*}

It is helpfull to write any octonion $a$ as%
\[
a=\sum_{l=0}^{l=7}u_{l}\mathbf{e}_{l}\text{, with }\mathbf{e}_{0}%
=1,\mathbf{e}_{1}=\mathbf{i,e}_{2}=\mathbf{j,e}_{3}=\mathbf{k,e}%
_{4}=\mathbf{e,e}_{5}=\mathbf{ie,e}_{6}=\mathbf{je,e}_{7}=\mathbf{ke}%
\]

with the following multiplication table%
\[
\left(
\begin{array}
[c]{cccccccc}%
\mathbf{e}_{0} & \mathbf{e}_{1} & \mathbf{e}_{2} & \mathbf{e}_{3} &
\mathbf{e}_{4} & \mathbf{e}_{5} & \mathbf{e}_{6} & \mathbf{e}_{7}\\
\mathbf{e}_{1} & -\mathbf{e}_{0} & \mathbf{e}_{3} & -\mathbf{e}_{2} &
\mathbf{e}_{5} & -\mathbf{e}_{4} & -\mathbf{e}_{7} & \mathbf{e}_{6}\\
\mathbf{e}_{2} & -\mathbf{e}_{3} & -\mathbf{e}_{0} & \mathbf{e}_{1} &
\mathbf{e}_{6} & \mathbf{e}_{7} & -\mathbf{e}_{4} & -\mathbf{e}_{5}\\
\mathbf{e}_{3} & \mathbf{e}_{2} & -\mathbf{e}_{1} & -\mathbf{e}_{0} &
\mathbf{e}_{7} & -\mathbf{e}_{6} & \mathbf{e}_{5} & -\mathbf{e}_{4}\\
\mathbf{e}_{4} & -\mathbf{e}_{5} & -\mathbf{e}_{6} & -\mathbf{e}_{7} &
-\mathbf{e}_{0} & \mathbf{e}_{1} & \mathbf{e}_{2} & \mathbf{e}_{3}\\
\mathbf{e}_{5} & \mathbf{e}_{4} & -\mathbf{e}_{7} & \mathbf{e}_{6} &
-\mathbf{e}_{1} & -\mathbf{e}_{0} & -\mathbf{e}_{3} & \mathbf{e}_{2}\\
\mathbf{e}_{6} & \mathbf{e}_{7} & \mathbf{e}_{4} & -\mathbf{e}_{5} &
-\mathbf{e}_{2} & \mathbf{e}_{3} & -\mathbf{e}_{0} & -\mathbf{e}_{1}\\
\mathbf{e}_{7} & -\mathbf{e}_{6} & \mathbf{e}_{5} & \mathbf{e}_{4} &
-\mathbf{e}_{3} & -\mathbf{e}_{2} & \mathbf{e}_{1} & -\mathbf{e}_{0}%
\end{array}
\right)
\]

Note that other multiplication tables could be defined (see \cite{baez}). In
analogy with the quaternions scalar and vectorial parts, we can also write $a$
as%
\[
a=S(a)+\mathbf{V}(a),\;\text{with }S(a)=u_{0},\;\mathbf{V}(a)=\sum_{l=1}%
^{l=7}u_{l}\mathbf{e}_{l}\mathbf{,}%
\]

The conjugate of an octonion $a$ is $\overline{a}=S(a)-V(a)\mathbf{=}%
\overline{a^{\prime}}-a^{\prime\prime}\mathbf{e}$ and its squared norm reads
$N_{a}^{2}=a\overline{a}$

A (very) important difference between quaternions and octonions is that the
latter, besides being non commutative are also non associative.

\bigskip

\textbf{Acknowledgement and comments}

It is a pleasure to acknowledge several discussions with Rossen Dandoloff,
Karol Zyczkowski and Perola Milman.

The main content of this paper was presented at the Dresde "Topology in
Condensed Matter Physics" colloquium in june 2002. Also included is a
reference to a paper by B. A. Bernevig and H.D. Chen, who have independantly
done the analysis of the three-qubit case along parallel lines. It is a
pleasure to acknowledge several discussions with Andrei Bernevig in that
context. Informations on more recent work with P. Milman on the the SO(3)
topological phase has been added. Also not presented in Dresde was the above
suggestion to use, for the three-qubit case, a three parameter foliation based
on the three partial Bloch sphere radii.

\begin{center}

\end{center}

\end{document}